\documentclass[intlimits,twoside,a4paper]{article}

\usepackage{amsmath,amssymb}
\usepackage{graphicx}

\usepackage[T2A]{fontenc}
\usepackage[cp1251]{inputenc}
\usepackage[eqsecnum]{cmpj2}

\issue{2013}{16}{2}{23008}

\doinumber{10.5488/CMP.16.23008}

\title[Universality classes and critical phenomena in confined liquid
systems]%
{Universality classes and critical phenomena in confined liquid
systems}
\author[A.V.~Chalyi \textsl{et al.}]{A.V. Chalyi\refaddr{label1},
      L.A. Bulavin\refaddr{label2}, V.F. Chekhun\refaddr{label3}, K.A. Chalyy\refaddr{label1}, L.M. Chernenko\refaddr{label4}, A.M.~Vasilev\refaddr{label2}, E.V. Zaitseva\refaddr{label1}, G.V. Khrapijchyk\refaddr{label1}, A.V. Siverin\refaddr{label2}, M.V. Kovalenko\refaddr{label2} }

\authorcopyright{A.V. Chalyi, L.A. Bulavin, V.F. Chekhun, K.A. Chalyy, L.M. Chernenko, A.M. Vasilev, E.V. Zaitseva, \\G.V. Khrapijchyk, A.V. Siverin, M.V. Kovalenko, 2013}

\addresses{
\addr{label1}Bogomolets National Medical University, Kyiv, Ukraine
\addr{label2}Taras Shevchenko Kiev National University, Kyiv, Ukraine
\addr{label3}Kavetskii Institute of Experimental Pathology, Oncology and Radiobiology, \\National Academy of Sciences of Ukraine
\addr{label4}Chuiko Institute of Surface Chemistry, National Academy of Sciences of Ukraine}
\date{Received April 17, 2013, in final form May 17, 2013}
\begin{document}

\maketitle
\begin{abstract}
It is well known that the similar universal behavior of infinite-size (bulk) systems of different nature requires the same basic conditions: space dimensionality; number components of order parameter; the type (short- or long-range) of the intermolecular interaction; symmetry of the fluctuation part of thermodynamical potential. Basic conditions of similar universal behavior of confined systems needs the same supplementary conditions such as the number of monolayers for a system confinement; low crossover dimensionality, i.e., geometric form of
restricted volume; boundary conditions on limiting surfaces;
physical properties under consideration. This review paper is aimed at studying all these conditions of similar universal behavior for diffusion processes in confined liquid systems. Special
attention was paid to the effects of spatial dispersion and low
crossover dimensionality. This allowed us to receive receiving correct nonzero
expressions for the diffusion coefficient at the critical point and to take into account the specific geometric form of the confined liquid
volume. The problem of 3D$\Leftrightarrow$2D dimensional
crossover was analyzed. To receive a smooth crossover for critical
exponents, the Kawasaki-like approach from the theory of mode
coupling in critical dynamics was proposed. This ensured a good
agreement between data of computer experiment and theoretical
calculations of the size dependence of the critical temperature
$T_{\mathrm{c}}(H)$ of water in slitlike pores. The width of the quasi-elastic
scattering peak of slow neutrons near the structural phase
transition in the aquatic suspensions of plasmatic membranes
(mesostructures with the typical thickness up to 10~nm) was studied.
It was shown that the width of quasi-elastic peak of neutron
scattering decreases due to the process of cell proliferation,
i.e., with an increase of the membrane size (including the membrane
thickness). Thus, neutron studies could serve as an additional
diagnostic test for the process of tumor formation.

\pacs 05.70.Jk, 68.18.Jk, 68.35.Rh, 61.12.-q, 82.56.Lz
\end{abstract}

\section{Introduction}

Second half of the previous century, especially its last decades,
was guided by two great achievements of experimental and
theoretical physics: (i) the revolutionary discoveries in
nanotechnologies, (ii) solution of problem of the 2$^{\mathrm{nd}}$ order phase
transitions. It is now considered generally accepted that the
achievements of nanosciences (including nanoelectronics and
nanomedicine) will determine the character of the 21$^{\mathrm{th}}$ century. It
is difficult to overestimate the consequences of creation of modern
picture of physics of phase transitions and critical phenomena
(physics of cooperative processes) based on precise experiments and
profound ideas of scale invariance (scaling) and renormalization group
\cite{1,2,2_2} as well as of the method of collective variables \cite{3,4}.
Unification of these two directions in the development of physics, which
started 40 years ago following the formulation of the scaling hypothesis
for spatially limited systems \cite{5,6,7}, continues to raise an
increasing interest of researchers to the study of phase transitions
and critical phenomena in mesoscale systems. In resent years
this interest touches not only upon magnetics and liquid crystals but
also extends to low- and high- temperature liquid systems.

The ideas of isomorphism of critical phenomena and phase transitions
\cite{8,9} make it possible to generalize the properties of confined liquids to the
systems not only of physical but also of other nature.

 This review paper is aimed at studying the conditions of similar universal behavior in confined liquids. Its structure is as follows. Section~2 ``Discussion'' consists of three subsections. The $1^{\mathrm{st}}$ subsection is devoted to the universal behavior of bulk and finite-size systems of different nature. In the $2^{\mathrm{nd}}$ subsection we consider the diffusion processes in mesoscale liquid systems with taking into account the additional factors of universality classes for finite-size systems, especially the spatial dispersion and the geometric form of finite-size volumes under consideration. The $3^{\mathrm{rd}}$ subsection is devoted to our original studies of the dimensional crossover (or smooth transition) between the properties of the bulk 3D and finite-size (even 2D) systems. For this purpose we introduce the Kawasaki-like analytical expression for such a 3$D\Leftrightarrow$ 2D dimensional crossover and propose the theoretical background for the results of computer simulations with the interpolation formula for the effective critical exponent $\nu(H)$. And finally, in section~3, we use the methods of neutron optics to study the temperature and size dependence of the width $\triangle E$ of the quasielastic neutron scattering peak near the structural phase transition in the plasmatic membranes (mesostructures with the typical thickness up to 10~nm). It is shown that studies of the width $\triangle E$ of the quasielastic neutron scattering peak can be applied as an effective tool for the tumor growth diagnostics.

\section{Discussion}
 \subsection{Universality classes for infinite (bulk) and finite-size systems}
Let us first recall the important notion of universality classes. Basic conditions of the similar universal behaviour for infinite-size (bulk) systems of different nature are well-known \cite{1,2,2_2,8,9}: (i) space dimensionality; (ii) the number of components of order parameters; (iii) the type (short- or long-range) of the intermolecular interaction; (iv) symmetry of Hamiltonian (fluctuation part of the thermodynamic potential).
Similar universal behaviour for confined systems needs the following basic conditions in addition to four previous ones: (v) geometric factors (the number of monolayers) for system confinement; (vi) low crossover dimensionality defined by the shape of the restricted volume, (see below for a more detailed explanation); (vii) the type of boundary conditions; (viii) the physical properties under consideration \cite{5,6,7,10,11,12}.
These basic conditions of similar universal behavior of confined systems will be illustrated herein below.

 \subsection{Diffusion processes in mesoscale liquid systems}
Methods of the theory of phase transitions in the spatially limited
systems are used here to study the diffusion coefficient of water
molecules in cylindrical pores, as well as the effects of spatial
dispersion and low crossover dimensionality
(geometrical form) on the diffusion processes \cite{5,6,7,10,11,12,13,14,15,16}.

Taking into account the fundamentals of thermodynamics and statistical
physics of irreversible processes as well as the modern theory of
critical phenomena in liquid systems, one can write down the
coefficient of self-diffusion
\begin{equation}
D=(L_{\mathrm{R}}+L_{\mathrm{S}})(\partial \mu/\partial\rho)_{T}\,.
\end{equation}
Here, $L_{\mathrm{R}}$ and $L_{\mathrm{S}}$ are regular and singular parts of the kinetic
Onsager coefficient, the derivative  $(\partial \mu/\partial\rho)_{T}$
is proportional to the inverse value of the isothermal
compressibility of liquids. In accordance with the theory of dynamic
scaling, the singular part $L_{\mathrm{S}}$ of the Onsager coefficient behaves
as the characteristic correlation length of the order parameter
fluctuations (for liquids -- fluctuations of density), namely:
$L_{\mathrm{S}}=L_{\mathrm{S}}^0(\tau^*)^{-\nu}$ where  $L_{\mathrm{S}}^0$ is the amplitude, $\tau^*$
is the corresponding temperature variable for the systems with
restricted geometry, and $\nu\approx 0.63$  is the critical index. This value for the correlation length critical exponent  $\nu$ is taken because the bulk classical liquids belong to the 3D Ising model universality class.
According to the scaling theory, the derivative $(\partial
\mu/\partial\rho)_{T}=(\partial
\mu/\partial\rho)_{T}^0(\tau^*)^{\gamma}$. Here, $(\partial
\mu/\partial\rho)_{T}^0$ is the amplitude of inverse isothermal
compressibility, and $\gamma\approx 1.24$ is the isothermal compressibility critical index in the 3D Ising model universality class.
Finally, the coefficient of self-diffusion in the spatially-limited
liquid system is described by the following formula:
\begin{equation}
D=\left[L_{\mathrm{R}}+L_{\mathrm{S}}^0(\tau^*)^{-\nu}\right](\partial
\mu/\partial\rho)_{T}^0(\tau^*)^{\gamma}.
\end{equation}

In a general case, while describing the dynamic phenomena, the following three
regions exist depending on temperature ``distance'' to the phase
transition point \cite{17}:
\begin{enumerate}
\item {\emph{Dynamic fluctuation region}}, where singular parts of
kinetic Onsager coefficients substantially prevail over its regular parts
($L_{\mathrm{S}}\gg L_{\mathrm{R}}$);
\item {\emph{Dynamic crossover region}}, where both parts of the
kinetic Onsager coefficients are of the same order of magnitude
($L_{\mathrm{S}}\approx L_{\mathrm{R}}$);
\item {\emph{Dynamic regular region}}, where singular parts
of kinetic Onsager coefficients are substantially less than its
regular parts ($L_{\mathrm{S}} \ll L_{\mathrm{R}}$).
\end{enumerate}

An answer to a question, which of these regions realize in
experiments or in natural conditions, depends on the value of
Ginzburg-Levanyuk number $\mathrm{Gi}$, which permits to estimate the role of
fluctuation effects. For weak aquatic solutions with $\mathrm{Gi}\approx0.3$ only
dynamic crossover and regular regions are expected to be observed
in reality.

It appears that the temperature variable $\tau^*$  is characterized
by the following formula in liquids with confined geometry
\cite{10,15,16}:
\begin{equation}
\tau^*=(G/S)^{\frac{1}{\nu}}+\left[1+(G/S)^{\frac{1}{\nu}}\right](\xi^*)^{-\frac
{1}{\nu}}\,.
\end{equation}

Here, $G$ is a geometrical factor which depends on the low crossover
dimensionality (geometrical form) of liquid volume [for the
plane-parallel layer $G=\pi$, while for cylindrical sample
$G=2.4048$ is the first zero of the Bessel function $J_0(z)$;
$S = L/a_0$ is the number of monolayers ($L$ is a linear size of the
system) in the direction of its spatial limitation, $a_0$ is the average
diameter of a molecule], $\xi^*=\xi/\xi_0$ is the dimensionless
correlation length of density fluctuations ($\xi_0$ is the amplitude
of correlation length which has the same order of magnitude as
$a_0$).

The size dependence of the self-diffusion coefficient $D(S)$ is theoretically estimated in the dynamic crossover region in accordance with formulae (2.2), (2.3) in \cite{14} (figures~1, 2). Obviously, in a general case of restricted systems for which
inequality  $\xi\gg L$ is correct, the first term
$(G/S)^{\frac{1}{\nu}}$ will prevail in (2.3), that is why the diffusion coefficient will
decrease at the fixed temperature while the linear sizes of a
system increase. In the opposite case, i.e.,  for a relatively large
linear size in the sense of inequality $\xi\ll L$ , the multiplier
$(\xi^*)^{-\frac {1}{\nu}}$ in the second term in (2.3) will play a
greater role. That is why the diffusion coefficient $D$ will grow
and will asymptotically approach the value of $D_0$ in the spatially
unlimited volume.
\begin{figure}[ht]
\centerline{\includegraphics[width=9cm,height=4cm]{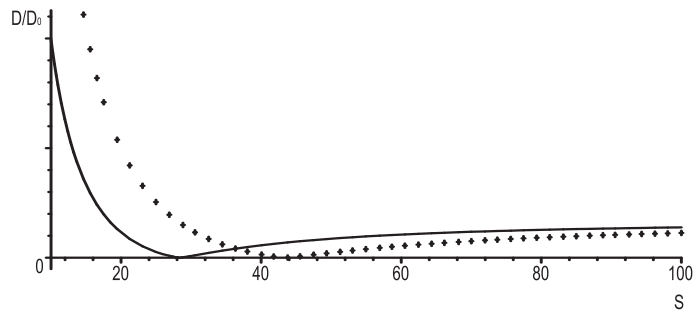}}
\caption{Dependence of self-diffusion coefficient $D$
on $S$.}
\end{figure}

\begin{figure}[ht]
\centerline{\includegraphics[width=9cm]{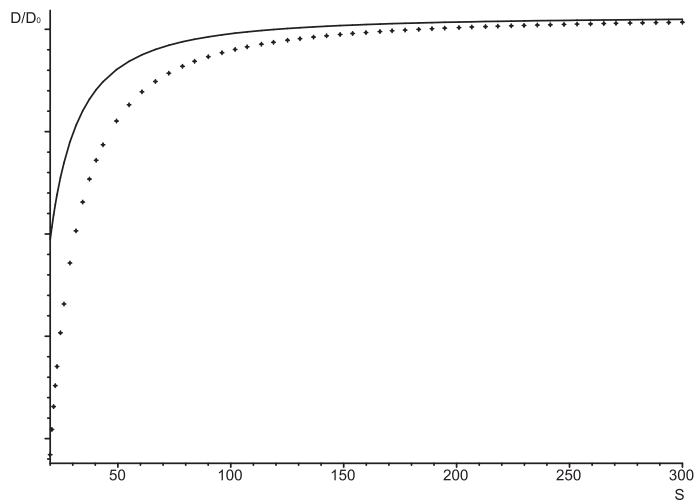}}
\caption{Size dependence of self-diffusion coefficient
at $S\gg \xi^*$.}
\end{figure}

Figure~1 illustrates a non-monotonous dependence of self-diffusion
coefficient $D$ (in relative units) on the parameter $S$ (the curve with
circles -- for the liquid system with cylindrical geometry, dotted
curve -- for geometry of plane-parallel layer), at the fixed
temperature deviation  $\tau = (T-T_{\mathrm{c}})/T_{\mathrm{c}}=-0.01$, where $T_{\mathrm{c}}$
is the critical temperature of a bulk phase.

An increase of the self-diffusion coefficient $D$ with the growth of $S$
(see right-hand  part of figure~1 at $S > 40$, and figure~2), is confirmed
both by experimental data \cite{18} in cylindrical pores within the range interval
of radius 40--150~nm and by theoretical calculations \cite{12} for the
self-diffusion coefficient of water molecules.

\begin{table}[!h]
\caption{Correspondence between the real system geometry and low crossover dimensionality $d_{\mathrm{LCD}}$.}
\begin{center}
\begin{tabular}{|c|c|c|}
\hline Real 3-dimensional confined &
A corresponding&
Low crossover\\
systems&borderline case&dimensionality\\
 \hline \hline  Plane-parallel layer, slitlike pore,
 & monomolecular
 &
 2\\
plane interphase, membrane, & plane&\\
synaptic cleft&& \\
 \hline Cylindrical pore, long pore with   &
 monomolecular
 &
1 \\
square or rectangular sections,& filament (line)&\\
 ionic channel&&\\
\hline  Sphere, cube, parallelepiped,
 &
 Point
(one molecule)
 &
 0 \\
ellipsoid of rotation, vesicle&&\\
 \hline
\end{tabular}
\end{center}
\end{table}
Theoretical studies performed in \cite{14} demonstrate the dependence of the self-diffusion coefficient on the geometric form of a liquid system or, in other words, on its low crossover dimensionality $d_{\mathrm{LCD}}$. This is briefly summarized in table~1. The low crossover dimensionality $d_{\mathrm{LCD}}$ determines
the limited spatial dimension of geometric objects (2$^{\mathrm{nd}}$ column)
 towards which the real investigated system (1$^{\mathrm{st}}$ column) passes if
its linear size (sizes) in the direction (directions) of spatial
limitation converge to a minimum possible size, i.e., to the molecule
diameter.  It is clear that a three-dimensional plane-parallel
layer transfers to the monomolecular plane (essentially -- 2D
object), while a three-dimensional cylindrical pore passes to the
monomolecular filament (essentially -- 1D object), while spheres or
cubes restricted at three directions have as its limit only one
molecule, i.e., 0D object. The last $3^{\mathrm{rd}}$ column of table~1 contains
the value of $d_{\mathrm{LCD}}$ for real spatially limited systems.

The analysis of the dependence of the self-diffusion coefficient on
$d_{\mathrm{LCD}}$ makes it possible to formulate the following conclusion valid for other
equilibrium and non-equilibrium properties of nano- and mesoscale
systems: with an increase of $d_{\mathrm{LCD}}$,  physical properties
of the spatially limited systems tend to their bulk values.

\pagebreak

Another conclusion concerns the temperature position
$\tau_{\mathrm{M}}(S)=[T_{\mathrm{c}}(S)-T_{\mathrm{c}}]/T_{\mathrm{c}}$ of the extremum (precisely -- minimum) of the
self-diffusion coefficient depending on the parameter $S$, i.e.,
linear size $L$ of the liquid systems with different crossover
dimensionalities.

\begin{figure}[!h]
\centerline{\includegraphics[width=0.95\textwidth]{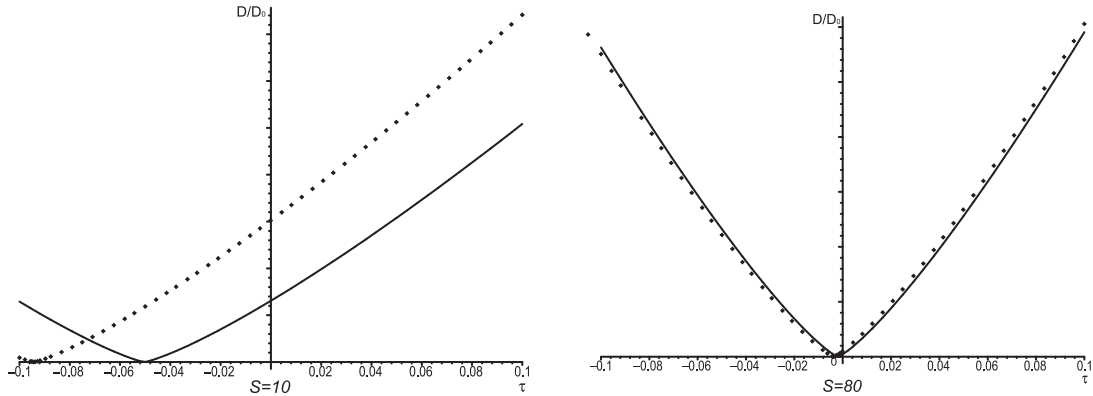}}
\caption{Dependence of the self-diffusion coefficient
$D$ on $\tau=(T-T_{\mathrm{c}})/T_{\mathrm{c}}$ at $S=\mathrm{const}$.}
\end{figure}
As it follows from figure~3 obtained in \cite{14}, the temperature deviation $\tau_{\mathrm{M}}(S)$ has a
negative value (in complete accordance with the scaling theory for spatially-limited systems); increases modulo at diminishing the $d_{\mathrm{LCD}}$; and tends to zero at increasing the linear sizes of the
system.

The above-mentioned results and conclusions are qualitatively confirmed by the data of the heat capacity temperature dependence in
confined liquids of different geometry \cite{19}. It follows that a heat
capacity maximum shifts to the region of lower temperatures and this
shift grows from a bulk phase to the liquids with $d_{\mathrm{LCD}}$ changing
from 2 to 1 and 0 (see table~1 where the real examples of such
spatially limited liquids are presented). \noindent

\begin{table}[!h]
\caption{Self-diffusion coefficient for water molecules and related properties (see text for the explanation).}
\vspace{2ex}
\begin{center}
\begin{tabular}{|c|c|c|c|c|c|c|c|}
\hline $\tau$  & $L_{\mathrm{R}}^*$&
$L_{\mathrm{S}}^*$&$L^*$&$(\partial\mu/\partial\rho)_{T}^*$&$ D^*$ &$ D_{\mathrm{water}}$,
$m^2/s$\\
\hline\hline $1.0$&$ 1$&$   10^{-3}$&$    1.001$ &$  1$&
$1.001$&$2.30\cdot10^{-9}$\\
 \hline $10^{-1}$&$ 1$&$ 4\cdot10^{-3}$&$ 1.004$&$
 5.6\cdot10^{-2}$&$5.622\cdot10^{-2}$&$ 1.29\cdot10^{-10}$\\

\hline $10^{-2}$&$    1$&$ 1.8\cdot10^{-2} $&$1.018$ &$
3.3\cdot10^{-3}$&$ 3.359\cdot10^{-3}$&$ 7.73\cdot10^{-12}$\\
\hline $10^{-3}$&$ 1$&$ 10^{-1}$&$ 1.1$&$ 1.9\cdot10^{-4}$&$    1.945\cdot10^{-4}$&$ 4.47\cdot10^{-13}$ \\
\hline $10^{-4}$&$ 1$&$ 3\cdot10^{-1}$&$ 1.3$&$ 1.1\cdot10^{-5}$&$ 1.43\cdot10^{-5}$&$   3.29\cdot10^{-14}$\\
\hline $10^{-5}$&$ 1$&$   1$&$   2.0$&$ 6.3\cdot10^{-7}$&$ 1.26\cdot10^{-6}$&$ 2.90\cdot10^{-15}$ \\
\hline $10^{-6} $&$   1$&$ 6.3$&$ 7.3$&$ 3.6\cdot10^{-8}$&$ 2.63\cdot10^{-7}$&$ 6.05\cdot10^{-16}$\\

\hline  $10^{-7}$&$ 1$&$ 25$&$ 26$&$ 2.1\cdot10^{-9}$&$ 5.46\cdot10^{-8}$&$ 1.26\cdot10^{-16}$\\
\hline
\end{tabular}
\end{center}
\end{table}
Analytical formulae obtained for the self-diffusion coefficient
$D^*=D/D_0$ make it possible to conduct numeral calculations of the
self-diffusion coefficient $D$ of a certain liquid almost in the whole
critical area $(0\leqslant  \tau \leqslant  1)$ using only a single parameter, i.e.,
the known value of amplitude $D_0
=L_{\mathrm{R}}^0(\partial\mu/\partial\rho)_{T}^0$. As an example, the results of
self-diffusion coefficient for water molecules are presented in
table~2. The value of the amplitude of self-diffusion coefficient for
water molecules  $D_0$ used for this purpose was  found in a regular
area far from the critical point:   $D_0=2.3\cdot 10^{-9}$~m$^2$/s.
This value was obtained experimentally for the molecules of water at
the temperature of $T = 293$~K, which corresponds to the temperature
deviation $\tau\approx -0.5$  from the critical temperature of water
of $T_{\mathrm{c}}=647$~K. Then, taking into account the value of $D_0$, as well as data for $D^*$ from the next to the last column of table~2, we
get numerical results for temperature dependence of the diffusion
coefficient of water molecules in the critical region (see the last
column of table~2).

The effects of spatial dispersion (nonlocality) being neglected results in
the physical properties in the critical points or the points of the 2nd
order phase transitions tending to infinity (i.e., isothermal compressibility,
magnetic susceptibility, isobar and isochoric heat capacities and
others) or to zero (i.e., coefficients of diffusion and thermal
diffusivity, speed of sound
and others). To take into account the effects of spatial
dispersion, the following idea is used \cite{20}: spatial dispersion
terms must be added to the values which become equal to zero in the
critical point (for example, added to the coefficient of diffusion of $D$ or
to the reverse value of the isothermal compressibility). Thus,
for the self-diffusion coefficient  one has the following formula:
\begin{equation}
 D^*(\xi ^*,S,k)=\frac{\left\{(G/S)^{\frac{1}{\nu}}+\left[1+(G/S)^{\frac{1}{\nu}}\right](\xi^*)^{-\frac{1}{\nu} }\mathrm{sign} \tau\right\}^{\gamma}+Bk^2}{\left\{(G/S)^{\frac{1}{\nu}}+\left[1+(G/S)^{\frac{1}{\nu}}\right](\xi^*)^{-\frac{1}{\nu} }\mathrm{sign} \tau\right\}^{\nu}+bk^2}\,.
\end{equation}

It follows from (2.4) that a minimum nonzero value of the diffusion
coefficient is equal to
\begin{equation}
D_{\mathrm{min}}^*(L)=\frac{(G/S)^{\frac{\gamma}{\nu}}+4\pi
^2B/L^2}{(G/S)^{\frac{\gamma}{\nu}}+4\pi ^2b/L^2}\,.
\end{equation}

 The same approach could be also used for taking into account the effects of temporal (frequency) dispersion of physical properties in the critical region. However, this problem will not be examined herein.

In the critical point ($\tau=0$), the self-diffusion coefficient of
the bulk phase is constant and nonzero:
$D=L_{\mathrm{R}}^0(\partial\mu/\partial\rho)_{T}^0B/b={\mathrm{const}}$ where $B$ and $b$
are coefficients of nonlocality. Figure~4 (solid curve)
illustrates this result. Two other graphs demonstrate temperature
dependence of the self-diffusion coefficient for liquids in
plane-parallel (dotted curve) and cylindrical (curve with circles)
confined geometry.
\begin{figure}[ht]
\centerline{\includegraphics[width=9cm]{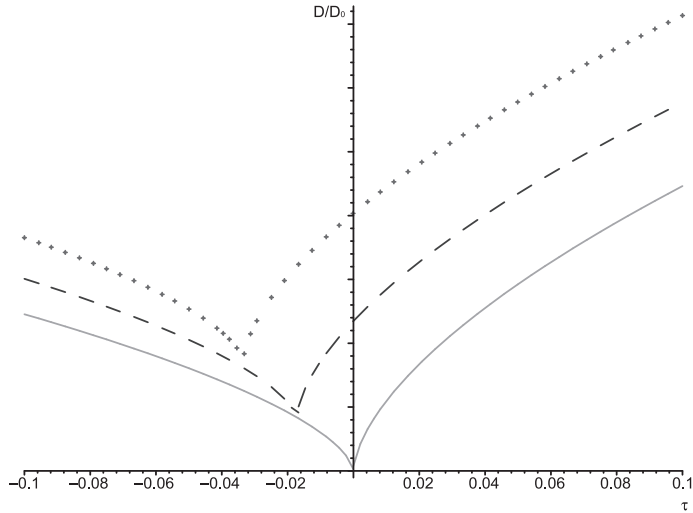}}
\caption{Temperature dependence of the self-diffusion
coefficient.}
\end{figure}

An important peculiarity of the self-diffusion coefficient was also
taken into account. Namely, its asymmetry (see figure~4), as it
follows from the inequality  $D_+^*>D_-^*$, where
$D_+^*=D(|\tau|,L,k)/D_0^+$ and $D_-^*=D(-|\tau|,L,k)/D_0^-$. There
are two reasons for such an inequality $D_+^*>D_-^*$: (i) change of
sign of temperature deviation $\tau$ in expressions for
$D(\pm|\tau|,L,k)$; (ii) inequality $D_0^+\neq D_0^-$ of
self-diffusion amplitudes in overcritical ($T>T_{\mathrm{c}}$) and subcritical
($T<T_{\mathrm{c}}$) regions. The same temperature dependence of the
self-diffusion coefficient (confirmed by independent theoretical
calculations \cite{21}) should be expected in experimental studies of
diffusion processes in finite-size liquids.

 \subsection{Dimensional crossover in finite-size liquid systems}

In this section we would like to pay attention to the following
problem: how the results of 3D systems can be transferred to the results of 2D
systems and vice versa. Of course, this transition cannot be very
sharp; it should be smooth and without discontinuities, i.e.,
crossover-like. Let us call this 3D${}\Leftrightarrow{}$2D transition
as the dimensional crossover. To describe the dimensional crossover
we shall take into account (i) an obvious fact that the critical
exponents in 3D and 2D systems have quite different numerical
values (see table~3); (ii) the results of computer experiments \cite{22}.
\begin{table}[h]
\caption{Values of the critical exponents in 2D and 3D systems \cite{9}
(* Ornstein-Zernike approximation).}
\begin{center}
\begin{tabular}{|c|c|c|c|c|c|c|c|}
\hline Space &&&&&&&\\
dimen&Theory or &$\alpha$&$\beta$&$\gamma$&$\delta$&$\nu$&$\eta$\\
siona&experiment&&&&&&\\
lity&&&&&&&\\
\hline
\hline
$ $  &  Landau &  $0$& $1/2 $&  $1$& $3$& $1/2^*$&$0^*$\\

\hline

2D& Ising model&  $+0 (ln|\tau|)$&    $1/8$&  $ 7/4$ &  $15$&   $1$&$1/4$\\
\hline

  3D&    Ising model &   $0.125   $ &  $ 0.3125   $&$   1.250$&$ 5  $ &$0.638 $& $0.041 $ \\
\hline
 3D&RG &$0.110$&$0.325$&$1.241$&$4.8$&$0.63$&$0.031$\\
\hline
 3D&    Experimental &  $0.11\pm$&    $0.33\pm$ &$1.23\pm$&   $ 4.6\pm$& $0.63\pm$& $0.04\pm$\\
&     data  &  $0.01$&    $0.01$ &$0.02$&   $ 0.2$& $0.01$& $0.02$\\
\hline
\end{tabular}
\end{center}
\end{table}

Let us consider a confined liquid system with, say, the geometry of
a plane-parallel layer. While reducing its width $L$ [or the number of
monolayers $S$, see formulae (3)--(5) in our approach], the system
will transfer from 3D to 2D geometry. This transition should
result in the change of critical exponents of classical liquids which
belong to the universality class of Ising model. The critical index
$\nu$ will shift its value from 0.63 to 1.0, the critical index
$\gamma$ -- from 1.24 to 1.75, etc. (table~3, \cite{9}).

To receive a smooth transition between two fixed quantities we would like to use the idea of Kawasaki  from the theory of mode coupling
\cite{23}. It permits to receive the so-called \emph{Kawasaki-like} formula for the critical exponents
inside 3D${}\Leftrightarrow{}$2D dimensional crossover:
\begin{equation}
y=n_3+\left[\frac{2}{\pi} \arctan(ax-b)-1\right]\frac{n_3-n_2}{2}\,.
\end{equation}
Equation~(2.6) provides an interpolation for any effective critical
exponent $y$ between its 3D and 2D values ($n_3$ and $n_2$,
respectively).
Here, $x=L/L_0$  is the dimensionless width of the
plane-parallel layer;
$L_0$ is the linear size of the system in
restricted geometry at which the crossover occurs (authors \cite{22}
consider  $L_0\approx 2.4$~nm for the slitlike pore);
\begin{figure}[!h]
\centerline{\includegraphics[width=9cm]{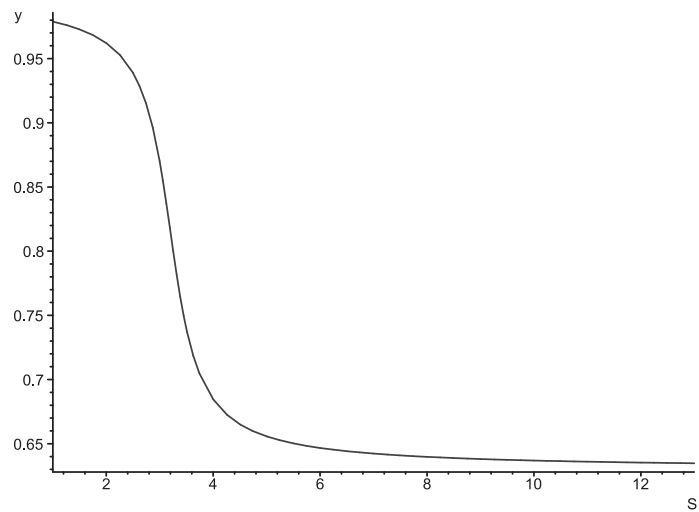}}
\caption{Dependence of the critical
exponent $\nu$ on $S$.}
\end{figure}
and $a$, $b$  are the
dimensionless parameters characterizing the slope and position of the
3D${}\Leftrightarrow{}$2D  crossover along the $S$-axis. Figure~5 provides the theoretical dependence of the effective critical
exponent $\nu$  on the number of monolayers $S$ in accordance with
equation (2.6). Parameters $a = 20$ and $b = 8$ were chosen to fit the
condition that the limiting 2D value of the critical exponent
$\nu=1$ corresponds to a system containing approximately one
monolayer.

The computer simulation experiment \cite{22} demonstrates the dependence of the
dimensionless pore critical temperature  $T_{\mathrm{c}}^{\mathrm{pore}}/T_{\mathrm{3D}}$ on the
pore size, i.e., the thickness $H$ of a slitlike pore or radius $R$ of a
cylindrical pore (figure~6). Closed squares and open circles
correspond, respectively, to slitlike and cylindrical pores filled with
water molecules. Dashed lines show the critical temperatures of the
bulk 3D water (upper line) and the 2D water (lower line). The
lowest square corresponds to the critical temperature of nearly 2D
water in slitlike pore with its thickness $H = 0.5$~nm. This value
of thickness $H$ refers to nearly one monolayer plane with taking
into account that the diameter of water molecule is equal to
$d\approx$0.3~nm.
\begin{figure}[ht]
\centerline{\includegraphics[width=7cm]{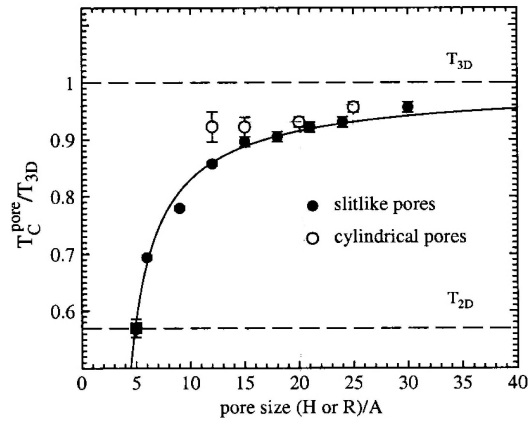}}
\caption{Size dependence of the
pore critical temperature (computer experiment \cite{22}).}
\end{figure}
\begin{figure}[ht]
\centerline{\includegraphics[width=9cm]{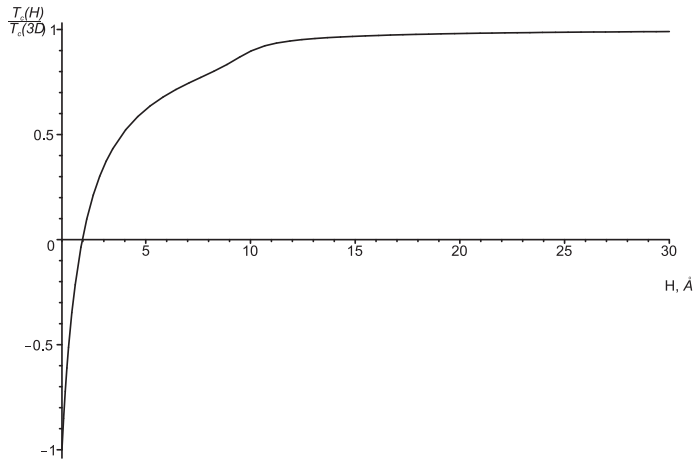}}
\caption{Size dependence of the
critical temperature in slitlike pore [finite-size scaling + formula
(2.6) for~$\nu$].}
\end{figure}

The finite-size scaling theory \cite{5,6,7,10,15,24} provides the following
formula for the shift of the critical temperature
$T_{\mathrm{c}}^{\mathrm{pore}}=T_{\mathrm{c}}(H)$ in comparison with its bulk value
$T^{\mathrm{3D}}=T_{\mathrm{c}}(\infty)$:
\begin{equation}
 \tau^*=\left[T_{\mathrm{c}}(H)-T_{\mathrm{c}}(\infty)\right]/T_{\mathrm{c}}(\infty)\sim H^{-\frac{1}{\nu}}.
 \end{equation}

The quite equivalent formula is as follows:
\begin{equation}
 \frac{T_{\mathrm{c}}(H)}{T_{\mathrm{c}}(\infty)}=1+kH^{-\frac{1}{\nu}}\,,
 \end{equation}
where $k$ is the coefficient of proportionality.

In order to check our interpolation formula (2.6) using the results of
computer experiment \cite{22},we substituted the size dependence of the
critical exponent $\nu (H)$  [see (2.6) and figure~5] into the formula
(2.8). It yields the size dependence of the critical temperature
$T_{\mathrm{c}}(H)$ in slitlike pores shown by figure~7. The agreement between
the computer experiment data and theoretical calculations seems to be
quite good. An additional curvature near the $3^{\mathrm{rd}}$ left-hand point
(figure~6) is even observed in the theoretical dependence $T_{\mathrm{c}}(H)$ in
figure~7.

It is interesting to stress that the beginning of the dimensional
crossover from 3D to 2D critical behavior takes place at the
slitlike pore thickness $H^{\mathrm{cross}}\approx 2.4$~nm. This value of
$H^{\mathrm{cross}}$ is mentioned in \cite{22} and can be observed in figure~7. It
corresponds to approximately 8 monolayers of water molecules in
a slitlike pore.

\section{Neutron studies and its medical applications}

The method of quasi-elastic neutron scattering (QENS) is a
powerful physical method of studying the dynamic properties of liquids
and aquatic suspensions of membrane mesostructures \cite{18}. In
particular, the direct relationship between the change of the
diffusion coefficient $D$ of water molecules and the sensitivity
of biological cells to antitumor drugs was examined based on the
theoretical calculations and precise QENS experimental information
in \cite{25,26}.

The corresponding theoretical background includes both contributions
of the collective and single-particle diffusion of water molecules.
The width of the quasi-elastic peak of slow neutron scattering
$\triangle E(q^2)$ can be presented by the following formula \cite{18,25,26}:
\begin{equation}
 \triangle E(q^2)=2\hbar
 D^{\mathrm{coll}}q^2+\frac{2\hbar}{\tau_0}\left[ 1-\frac{\exp(-2W)}{1+(D-D^{\mathrm{coll}})q^2\tau_0^{-1}}\right].
 \end{equation}
Here, $D^{\mathrm{coll}}$ is the collective contribution to the diffusion
coefficient of water molecules, $D$ is the diffusion coefficient
with collective and single-particle contributions, $W$ is the energy
of activation (Debay-Waller factor), $q=(4\pi/\lambda) \sin{\theta/2}$
is the change of neutron wave vector, $\theta$ is the scattering
angle of a neutron beam, $\tau_0\approx 10^{-10}\div 10^{-12}$ s is
the mean lifetime of hydrogen bonds.

Thus, the width of the neutron quasi-elastic scattering peak
$\bigtriangleup E(q^2)$ makes it possible to calculate $D$  and
$\tau_0$, i.e., the dynamic characteristics of interaction between
water molecules and its environment.

Neutron analysis of the width $\triangle E(q^2,\tau)$ in aquatic
solutions of plasmatic membranes of tumor cells is a promising
biomedical direction of studies near structural phase transitions
such as the cell proliferation \cite{27}. It is known (see, for example,
\cite{28,29,30}) that the mobility of polar groups as well as rotational
mobility of carbohydrate chains changes near the phase
transitions of cell structures. Cooperative processes in the membranes
which are isomorphous to phase transitions in liquid mixtures play
an important role in the mechanisms of ionic transport, amplification of
external stimuli, diffusion processes in membrane memory, etc.

The plasmatic membranes of cells are typical mesostructures with a
characteristic thickness about 10~nm. Therefore, in accordance with
the hypothesis of dynamic scaling and expressions for kinetic
coefficients in spatially limited liquids (see the previous results in
this paper), we may write the following formula for the width of
quasi-elastic peak of slow neutron scattering in slitlike pores with
its thickness~$H$:
\begin{equation}
 \triangle E(q^2,\tau)=\triangle E_0(q^2)\left\{(G/S)^{\frac{1}{\nu}}+\left[1+(G/S)^{\frac{1}{\nu}}\right](\xi^*)^{-\frac{1}{\nu}}\mathrm{sign} \tau \right\}^{\gamma}.
 \end{equation}
 Here, $\triangle E_0(q^2)=2\hbar D_0q^2$ and $D_0$ are the amplitudes of the width of quasi-elastic peak and diffusion coefficient.

The main theoretical result of equations~(3.2) is as follows: the width
$\triangle E$ of quasi-elastic peak of slow neutrons scattering narrows (as well as the diffusion coefficient $D$ decreases),
while the process of proliferation with increasing $S\sim H$ takes
place in the mesostructure of plasmatic membranes.

Such a size dependence of width  $\triangle E$  in bulk aquatic suspensions of plasmatic membranes for the case $H/\xi\leqslant  1$  can be explained as follows. If $H > H_{\mathrm{D}}$, i.e., for sizes $H$ larger than the characteristic size $H_{\mathrm{D}}$  at which the dynamic crossover region is realized \cite{17,31}, one has the following expression:   $\triangle E (q^2,\tau)/\triangle E_0(q^2)\sim H^{-\gamma/\nu}\approx H^{-2}$, while  $\gamma/\nu=2-\eta\approx 1.96$. In the fluctuation region ($H < H_{\mathrm{D}}$), at which singular parts of the kinetic Onsager coefficients should be taken into account, the size dependence of  $\triangle E$ becomes smoother: $\triangle E(q^2,\tau)/\triangle E_0(q^2)\sim H^{(-\gamma+\nu)/\nu}\approx H^{-1}$.

 Thus, the studies of the width of quasi-elastic peak of slow neutron
scattering depending on the thickness of the membrane mesostructures, which
changes in the process of the cell proliferation, can serve as an
additional diagnostic test for the process of tumor formation.

\section{Conclusion}
In this review paper, we have investigated the specific features of mesoscale liquids in the critical region. We have shown that, having taken into account the actual factors of a liquid system at restricted geometry, such as number monolayers in confined systems, low crossover dimensionality, etc., the hypothesis of the universality may be essentially generalized for the finite-size systems of different nature.
There is another important problem discussed in this paper, namely, the 3$D\Leftrightarrow$ 2D dimensional crossover. We have proposed the interpolation formula (2.6) to get a smooth transition from 3D to 2D values of the critical exponent $\nu$  confirming the results of computer experiments.
We also hope that the further development of the physics of the 1st and continuous the 2nd order phase transitions will make a great contribution to biomedical applications; especially it will help to formulate new ideas and methods of diagnostics and to prevent the process of tumor formation.

\newpage
\ukrainianpart

\title{Класи універсальності та критичні явища
в обмежених рідинних системах}

\author{О.В.~Чалий\refaddr{label1}, Л.О.~Булавін\refaddr{label2}, В.Ф.~Чехун\refaddr{label3},   К.О.~Чалий\refaddr{label1}, Л.М.~Черненко\refaddr{label4}, О.М.~Васильєв\refaddr{label2}, О.В.~Зайцева\refaddr{label1}, Г.В.~Храпійчук\refaddr{label1}, О.В.~Северин\refaddr{label2}, М.В.~Коваленко\refaddr{label2}}

\addresses{
\addr{label1}  Національний медичний університет імені О.О.Богомольця
\addr{label2}Київський національний університет імені Тараса Шевченка
\addr{label3}Інститут експериментальної патології, онкології та радіології імені Кавецького НАН України
\addr{label4}Інститут хімії поверхні імені О.О.Чуйко НАН України}

\makeukrtitle

\begin{abstract}
Подібність універсальної поведінки систем великих розмірів різної природи вимагає однаковості таких основних умов: вимірності простору, числа компонент параметра порядку; коротко- або далекодіючих міжмолекулярних взаємодій; симетрії флуктуаційної частини термодинамічного потенціалу. Основні умови подібності універсальної поведінки для просторово обмежених систем доповнюються однаковими додатковими умовами: кількістю моношарів у напрямку просторового обмеження системи; нижньою кросоверною вимірністю, тобто геометричною формою обмеженого об'му; граничними умовами на обмежуючих поверхнях; фізичними властивостями, які розглядаються. Метою цієї оглядової статті було вивчення умов подібності універсальної поведінки процесів дифузії у просторово обмежених рідинних системах. Особливу увагу було приділено ефектам просторової дисперсії і нижньої кросоверної вимірності. Це дозволило отримати правильні ненульові вирази для коефіцієнта дифузії у критичній точці з урахуванням конкретної геометричної форми обмеженого об'єму рідини. При розгляді проблеми 3$D\Leftrightarrow$ 2D вимірного кросовера були отримані оригінальні результати для плавного переходу критичних індексів за допомогою підходу, схожого на метод Кавасакі в теорії динамічного скейлінгу. Це призвело до гарного узгодження між даними комп'ютерного експерименту  і теоретичними розрахунками залежності величини критичної температури $T_{\mathrm{c}}(H)$ води від товщини щілиноподібних пор. Було досліджено ширину квазіпружного піку розсіяння повільних нейтронів поблизу структурного фазового переходу в водних суспензіях плазматичних мембран (мезоструктур з типовою товщиною до 10 нм). Доведено, що ширина квазіпружного піку розсіяння нейтронів повинна зменшитися внаслідок процесу клітинної проліферації, тобто із збільшенням розміру мембрани (у тому числі товщин мембран). Таким чином, нейтронні дослідження можуть слугувати додатковим діагностичним тестом для виявлення процесу утворення пухлини.
\keywords класи універсальності, обмежені рідинні системи, просторова дисперсія, нижня кросоверна вимірність, вимірний кросовер, ширина квазіпружного піку, нейтронне розсіяння
\end{abstract}
\end{document}